\documentclass[prx,notitlepage,twocolumn,superscriptaddress]{revtex4-1} 
\usepackage{amssymb}
\usepackage[charter]{mathdesign}
\usepackage{bbold}
\usepackage{hyperref}  
\hypersetup{plainpages=false,colorlinks=true,linkcolor=blue, citecolor=blue, urlcolor=blue}

\usepackage{graphicx}
\usepackage{amsmath}
\usepackage{color}
\usepackage{float} 
\usepackage{multirow}
\usepackage{placeins}
\usepackage{xcolor}
\pagecolor{white}

\begin{document}

\title{Dynamical Variational Monte Carlo as a quantum impurity solver: Application to Cluster Dynamical Mean-Field Theory}
\author{P. Rosenberg}
\affiliation{D\'epartement de Physique, RQMP \& Institut Quantique, Universit\'e de Sherbrooke, Qu\'ebec, Canada J1K 2R1}
\author{D. S\'en\'echal}
\affiliation{D\'epartement de Physique, RQMP \& Institut Quantique, Universit\'e de Sherbrooke, Qu\'ebec, Canada J1K 2R1}
\author{A.-M. S. Tremblay}
\affiliation{D\'epartement de Physique, RQMP \& Institut Quantique, Universit\'e de Sherbrooke, Qu\'ebec, Canada J1K 2R1}
\author{M. Charlebois}
\affiliation{D\'epartement de Physique, RQMP \& Institut Quantique, Universit\'e de Sherbrooke, Qu\'ebec, Canada J1K 2R1}
\affiliation{D\'epartement de Chimie, Biochimie et Physique, Institut de Recherche sur l’Hydrog\`ene, Universit\'e du Qu\'ebec \`a Trois-Rivi\`eres, Trois-Rivi\`eres, Qu\'ebec G9A 5H7, Canada}

\begin{abstract}
Two of the primary sources of error in the Cluster dynamical mean-field theory (CDMFT) technique arise from the use of finite size clusters and
finite size baths, which makes the development of impurity solvers that can treat larger systems an essential goal. In this work we introduce an impurity solver based on the recently developed dynamical variational Monte Carlo (dVMC) method. Variational Monte Carlo possesses a favorable scaling as a function of system size, which enables the treatment of systems beyond the reach of current exact diagonalization solvers. To benchmark the technique, we perform a systematic set of CDMFT calculations on the one-dimensional Hubbard model. We compare to results obtained with an exact diagonalization solver for small clusters, and against the exact solution in the thermodynamic limit obtained by Lieb and Wu \cite{Lieb-Wu} for larger clusters. The development of improved impurity solvers will help extend the reach of quantum cluster methods, which can be applied to a wide range of strongly-correlated electron systems, promising new insights on their emergent behavior.
\end{abstract}

\maketitle

\section{Introduction}
Strongly-correlated many-electron systems are one of the central challenges of condensed matter physics.
These systems generally do not permit exact solutions, expect for selected limiting cases. The lack of
robust quantitative descriptions of these systems necessitates
the development of ever-improving approximative and numerical approaches. 
One set of computational
techniques that have proven quite successful in the treatment of strongly-correlated electrons are the quantum 
cluster methods, comprising Cluster Perturbation Theory (CPT)~\cite{gros_cluster_1993, senechal_spectral_2000,Senechal2002}, 
Cluster Dynamical Mean-Field Theory (CDMFT), and Dynamical Cluster Approximation (DCA) ~\cite{hettler_nonlocal_1998,hettler_dynamical_2000-1,aryanpour_analysis_2002,maier_quantum_2005-3,LTP:2006,lichtenstein_antiferromagnetism_2000-4, kotliar_cellular_2001-3,maier_quantum_2005-3}, among others. 

The principle underlying these techniques is to represent the system as a finite size cluster
embedded in an infinite lattice. The self-energy of the cluster can then be used to approximate
the self-energy of the infinite system. The effect of the infinite lattice on the cluster is incorporated
by adding additional terms to the cluster Hamiltonian, or, in the case of CDMFT, bath degrees of 
freedom whose values are determined self-consistently.

One of the central components of the CDMFT procedure is to solve the cluster-bath
problem. For discrete bath representations, most current impurity solvers are based on exact diagonalization, which
is limited to fairly small cluster sizes given the exponential scaling of the algorithm with the dimension of the Hilbert space.
This limitation leads to the two sources of systematic error in the CDMFT technique, finite size clusters
and finite size baths \cite{Senechal2010-2}. In the limit of infinite cluster size, the CDMFT result represents the thermodynamic limit, however,
for small clusters the finite size error can be significant. Unlike the case of an infinite cluster, an infinite bath connected to
a finite cluster does not represent the thermodynamic limit, but larger baths provide an improved representation of the effect
of the environment on the cluster. The development of impurity solvers that can treat larger clusters will help minimize these 
fundamental sources of error and improve the capability of this already powerful method.

In this work we introduce an impurity solver for CDMFT based on the dynamical variational Monte Carlo technique (dVMC) \cite{Charlebois2020}.
This approach is sign-problem free and scales polynomially with system size, which permits the treatment of cluster
sizes beyond the reach of current exact diagonalization solvers. These benefits however come at the cost of statistical errors
inherent to all Monte Carlo approaches and systematic errors due to the ansatz for the ground state and excitations. The method has 
already demonstrated impressive accuracy as an impurity solver in CPT calculations on the hole-doped two-dimensional Hubbard model 
\cite{Rosenberg2022}, and here we extend and apply the technique within CDMFT. To gauge the accuracy of the approach we perform a set 
of benchmarks on the one-dimensional Hubbard model. We compare our CDMFT-dVMC results to CDMFT-ED results on small to intermediate
size clusters before treating larger clusters, whose results are compared to the exact solution obtained by Lieb and Wu \cite{Lieb-Wu}.

We organize the remainder of the paper as follows.
We introduce the method in section~\ref{se::method}. In section~\ref{se::results}
we present a set of benchmarks, first for small and intermediate clusters, followed by
large clusters. Finally, we discuss the performance of the method and possible improvements in 
section~\ref{se::discussion}.  Details of the technique are provided in a set of Appendices. 

\section{Method}
\label{se::method}
The method we present in this work is based on the dVMC technique introduced in
Ref.~\cite{Charlebois2020}, which was designed to treat systems with periodic boundary
conditions and translational invariance. The technique was recently extended to treat systems 
with open boundary conditions, which enables its use
as an impurity solver in various quantum cluster methods \cite{Rosenberg2022}. In Ref.~\cite{Rosenberg2022}
the technique was applied within CPT, which does not include bath sites, nor
does it involve a self-consistency procedure. Here we implement the approach within
CDMFT, which includes both of these ingredients.

In this section, we present some relevant details of the generalized dVMC method, a more complete discussion can be found in Refs.~\cite{Charlebois2020,Rosenberg2022}. In the Supplemental Material \cite{supp_mat}, the reader can find the code implementing the algorithm described here, which was used to obtain the results presented below.


\subsection{Green function from generalized dVMC}
\label{sec:general_dVMC}

As in Refs.~\cite{Charlebois2020,Rosenberg2022}, the generalized dVMC technique uses variational Monte Carlo to optimize a ground state ansatz describing a system of $N_e$ electrons, which is then used to obtain the Green function.
Within the generalized dVMC approach, the Green function matrix at a complex frequency $z$ is computed according to~\cite{Rosenberg2022}:
\begin{align}
\mathbf{G}^{\pm}(z) &= 
\mathbf{S}((z\pm\Omega)\mathbb{1} \mp\mathbf{H})^{-1} \mathbf{S}
\label{eq:Gpm3}
\\
&= \mathbf{Q}((z\pm\Omega)\mathbb{1} \mp \mathbf{E})^{-1}\mathbf{Q}^\dag,
\label{eq:Gpm4}
\end{align}
%
%
where $\mathbf{S}$ is the overlap matrix of the non-orthogonal basis used to express the one particle excited sectors of the Hamiltonian operator in matrix form: $\mathbf{H}$.
$\mathbf{Q}\equiv \mathbf{S}^{1/2}\mathbf{U}$ and $\mathbf{U}$ and $\mathbf{E}$ are the eigenvectors and eigenvalues
respectively of the matrix $\mathbf{M} \equiv \mathbf{S}^{-1/2}\mathbf{H} \mathbf{S}^{-1/2}$.
To understand this formula intuitively, recall that the Green function is obtained from states with one more or one less particle compared with the ground state. 
These states are obtained~\cite{Charlebois2020,Rosenberg2022} from an algorithm that generates independent but non-orthogonal basis states whose overlap is represented by the matrix $\mathbf{S}$.  

The matrices $\mathbf{S}$ and $\mathbf{H}$ are of dimension $2 N N_{\rm exc}$, where $N$ is the number of sites and $ N_{\rm exc}$ is the number of single-particle excitations, i.e., the number of many-body states containing one more (fewer) electron than the ground state.  We select a physically motivated subset of all possible single-particle excitations based on locality ~\cite{Charlebois2020,Rosenberg2022}.

The factor of 2 reflects the presence of two spin species; in the case of one spin species, these matrices are of dimension $N N_{\rm exc}$.
The matrix elements of $\mathbf{S}$ and $\mathbf{H}$ are computed with respect to the variational ground state, obtained according to the algorithm described in Refs.~\cite{misawa_mvmcopen-source_2019-1,VMC2008,Charlebois2020,Rosenberg2022}.
The plus and minus sign of Eq.~\eqref{eq:Gpm3} refer to the electron and hole Green functions matrices respectively, and they have been omitted from the matrices $\mathbf{S}^\pm$, $\mathbf{H}^\pm$, etc. to remain concise.
 

As in Ref.~\cite{Rosenberg2022}, we apply a filtering algorithm to the overlap matrix $\mathbf{S}$ 
to reduce Monte Carlo noise and ensure that the matrix is positive definite. In Ref.~\cite{Rosenberg2022}, 
the filtering procedure removed only negative eigenvalues of the overlap matrix. Here we perform an additional filtering, which
removes the very small positive eigenvalues of the overlap matrix.
In order to determine the number of states to filter, we perform a singular value decomposition, $\mathbf{S}=\mathbf{U}_\textmd{\tiny{SVD}}\boldsymbol{\Sigma}_\textmd{\tiny{SVD}}\mathbf{V}^\dagger_\textmd{\tiny{SVD}}$, where the matrix $\boldsymbol{\Sigma}_\textmd{\tiny{SVD}} = \textmd{diag}(s_1,s_2,\hdots,s_{NN_\textmd{exc}})$ contains
the singular values of $\mathbf{S}$. Note that $\boldsymbol{\Sigma}_\textmd{\tiny{SVD}}$ is {\it{not}} related to the self-energy $\boldsymbol{\Sigma}(\omega)$ defined below. The smallest singular value we keep is given by $s_\textmd{min} = (s_\textmd{max}) \times 10^{-k}$, where $k$ is a condition number. We fix this condition number, for the sake of consistency, to $k=6$. Everything below $s_\textmd{min}$ is filtered as detailed in Ref.~\cite{Rosenberg2022}.

%

The Green function is then obtained by summing the electron and hole Green function matrices and keeping
only the sector corresponding to the trivial excitation, $m=n=0$, namely $c^{(\dag)} \vert \Omega \rangle $:
\begin{align}
G_{ij,\sigma}(z) = \left[ \mathbf{G^+}(z) + \mathbf{G^-}(z) \right]_{ij,\sigma=\sigma^\prime,m=n=0},
\end{align}
where the indices $i,j$ denote site numbers and $m,n$ denote excitation numbers. Let us underline here that the boldface font used up until this point included $i,j$ and $m,n$ but from now on, it will include only $i,j$. Indeed, the $m,n$ degrees of freedom are only necessary to sample the excited sectors of the Hamiltonian ($\mathbf{H}^\pm$), but not to express the cluster Green function $G_{ij,\sigma}(z)$.

\subsection{Cluster dynamical mean field theory (CDMFT)}

We implement the dVMC impurity solver within the CDMFT technique, so
in this section we present the essential components of the CDMFT approach.  
In CDMFT, the infinite lattice system is modeled by an impurity Hamiltonian
composed of a set of interacting cluster sites coupled to non-interacting bath sites.
The impurity Hamiltonian has the general form,

\begin{align}
\hat{H}_\textrm{imp} = &\sum_{ij,\sigma}^{N_c} (t_{ij} \hat{c}^\dagger_{i\sigma}\hat{c}_{j\sigma} + \text{h.c.}) + U\sum_i^{N_c} \hat{n}_{i\uparrow} \hat{n}_{i\downarrow}-\mu\sum_{i,\sigma}^{N_c}\hat{n}_{i\sigma} \notag\\
			&+\sum_{i\lambda,\sigma}(\theta_{i\lambda} \hat{c}^\dagger_{i\sigma}\hat{b}_{\lambda\sigma} + \text{h.c.})
			+\sum_{\lambda}^{N_b}\varepsilon_\lambda \hat{b}^\dagger_{\lambda\sigma}\hat{b}_{\lambda\sigma},
\label{eq:H_imp}
\end{align}
where $\hat{c}_{i\sigma}$ annihilates an electron of spin $\sigma =\,\, \uparrow,\downarrow$ on a cluster site $i = 1,\hdots,N_c$, and $\hat{b}_{\lambda\sigma}$
annihilates an electron of spin $\sigma$ on a bath site labeled by $\lambda = 1,\hdots, N_b$, where $N_c$ and $N_b$ are the number of cluster sites and bath sites,
respectively. The bath is parametrized by cluster-bath hybridization terms, $\theta_{i\lambda}$,
and bath site energies, $\varepsilon_\lambda$~\cite{Caffarel:1994,Koch2008,Senechal2010}. 
An illustration of the bath configuration used in this work is given in Fig.~\ref{fig:bath_configs}.

The cluster Green function, $\mathbf{G}_c(\omega)$,  is obtained by the impurity solver, which provides an
efficient means of extracting the self-energy $\boldsymbol{\Sigma}(\omega)$ at any complex frequency, $\omega$.
The Green function of the original lattice Hamiltonian can then be computed using the cluster self-energy and the
non-interacting lattice Green function, $\mathbf{G}_0(\mathbf{\tilde{k}},\omega)$,
\begin{equation}
\mathbf{G}^{-1}(\mathbf{\tilde{k}},\omega) = \mathbf{G}_0^{-1}(\mathbf{\tilde{k}},\omega)  - \boldsymbol{\Sigma}(\omega).
\label{eq:Glatt}
\end{equation}
The momentum $\mathbf{\tilde{k}}$ runs over the reduced Brillouin zone of the superlattice of clusters.

In order to compare the Green function of the lattice to that of the cluster, the lattice Green function is projected back onto the cluster,
\begin{equation}
\bar{\mathbf{G}}(\omega) = \frac{N_c}{N}\sum_{\mathbf{\tilde{k}}} \left[\mathbf{G}_0^{-1}(\mathbf{\tilde{k}},\omega)  - \boldsymbol{\Sigma}(\omega)\right]^{-1}.
\label{Gproj}
\end{equation}
The remainder of the CDMFT procedure consists in finding a set of bath parameters that makes the cluster Green function, $\mathbf{G}_c(\omega)$,
and the projected lattice Green function, $\bar{\mathbf{G}}(\omega)$, as close as possible to each other~\cite{kotliar_cellular_2001-3}. This can be done by minimizing a distance function~\cite{Caffarel:1994,Senechal2010}.
A new cluster Green function can then be computed with this new set of
bath parameters and the procedure is repeated
until convergence. In the case of the dVMC impurity solver, we determine convergence by inspection
of the bath parameters (see Appendix \ref{App:vs_Nb}).

After convergence, the lattice Green function $\mathbf{G}(\tilde{\mathbf{k}},\omega)$ (Eq.~\ref{eq:Glatt}) can then be used to compute the average value of observables.
The average lattice density is computed with:
\begin{equation}
n = \int_{-\infty}^0 d\omega \sum_{\tilde{\mathbf{k}}}\mathrm{Tr}[\mathbf{G}(\tilde{\mathbf{k}},\omega)].
\end{equation}
This will be the main focus of Sec.~\ref{se::results}


\section{Results}
\label{se::results}
\subsection{Model}
As an initial test of the CDMFT-dVMC technique, we perform a set of calculations on the one-dimensional
Hubbard model. This model has been solved exactly \cite{Lieb-Wu}, and has been well studied by other numerical techniques \cite{Koch2008,Bolech2003,Capone2004,Kyung2006,Balzer2008}, which makes it an ideal testbed to assess the accuracy of the approach.

The model has the following Hamiltonian,
\begin{equation}
\hat{H} = \sum_{\langle ij \rangle,\sigma} -t (\hat{c}^\dagger_{i\sigma}\hat{c}_{j\sigma} + \text{h.c.}) + U\sum_i \hat{n}_{i\uparrow} \hat{n}_{i\downarrow}-\mu\sum_{i,\sigma}\hat{n}_{i\sigma}
\end{equation}
where $t$ is the amplitude for nearest neighbor hopping, $U$ is the on-site interaction strength, and $\mu$ is the chemical potential. 
In the results that follow we take $U/t = 4$. 

\subsection{Benchmark against ED solver}

In this section we present a series of benchmark CDMFT calculations on small clusters, which provide a direct comparison of the performance of the dVMC impurity solver to that of an exact diagonalization solver. 

\begin{figure}[!ht]
    \begin{center}
           \includegraphics[width=0.5\columnwidth]{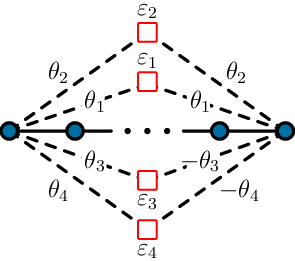}
    \end{center}
    \caption{Bath configuration. Bath sites couple to both edges of the cluster, with half of the bath
    sites belonging to the symmetric representation of the reflection symmetry, and the other half
    belonging to the anti-symmetric representation. Symmetry considerations are useful to define bath parameters~\cite{Koch2008,Liebsch:2009,PhysRevB.99.184510}. When more bath sites are used, the number of symmetric and anti-symmetric independent baths remains the same.
    \label{fig:bath_configs}
}
\end{figure}

We first study a cluster of two sites with four bath sites.
We chose a bath configuration in which each bath site is connected to the edge sites at either end of the cluster,
where two bath sites, with energies $\epsilon_1$ and $\epsilon_2$, belong to the symmetric representation of the reflection 
symmetry and the other bath sites, with energies $\epsilon_3$ and $\epsilon_4$, belong to the anti-symmetric representation.
The role of the bath sites is to represent the environment surrounding the cluster, therefore they couple only to the edges of the cluster.
The bath configuration is illustrated in Fig.~\ref{fig:bath_configs}. These small clusters can be
treated by exact diagonalization, which provides a reliable assessment of the accuracy of the dVMC solver.

\vspace{0.2cm}

\begin{figure}[!ht]
    \begin{center}
           \includegraphics[width=\columnwidth]{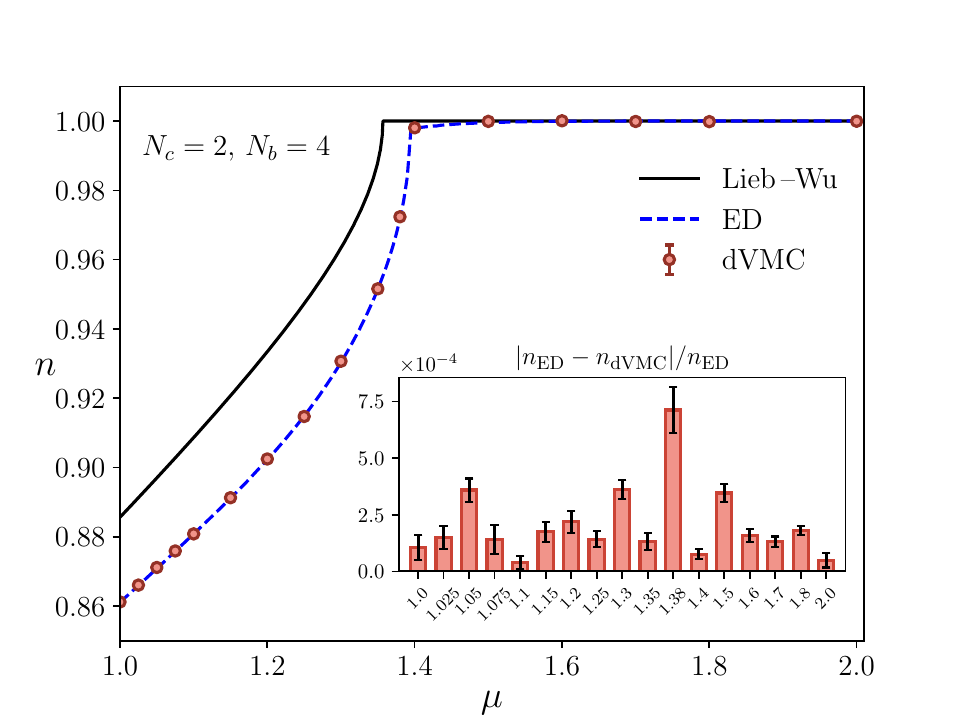}
    \end{center}
    \caption{Density versus chemical potential for a two site cluster with four bath sites.
    The inset shows the relative error between the result from the dVMC solver and the result from the ED solver
    at each value of $\mu$ studied. The error bars represent the error of the CDMFT-dVMC result 
    (see Appendix~\ref{App:vs_Nb} for details). Note that the same error is given in the inset as a fraction
    of the ED result. The exact result from Lieb \& Wu \cite{Lieb-Wu} is given by the black curve. \label{fig:L2_4b}
}
\end{figure}

In Fig.~\ref{fig:L2_4b} we show the average lattice density computed via CDMFT versus chemical potential. We observe that the results
obtained by the dVMC impurity solver (red circles) are in excellent agreement with the exact diagonalization results (ED, dashed blue line).
As shown in the inset, the relative error is below $0.1\%$ for the range of chemical potentials we have considered, which spans
from the metallic to the insulating state. This is an important initial demonstration of the capability of dVMC as a CDMFT impurity solver, indicating
that it can reproduce the results of CDMFT with an ED impurity solver on small clusters with quantitative accuracy.
Note that we also show the exact result obtained by Lieb and Wu for an infinite 1D interacting lattice from the Bethe ansatz \cite{Lieb-Wu} as a black line throughout this paper. It is only a reference for now, as in the best case scenario, we do not expect dVMC to obtain a result closer to the Lieb\,--Wu result than that obtained with ED.

\begin{figure}[!ht]
    \begin{center}
           \includegraphics[width=\columnwidth]{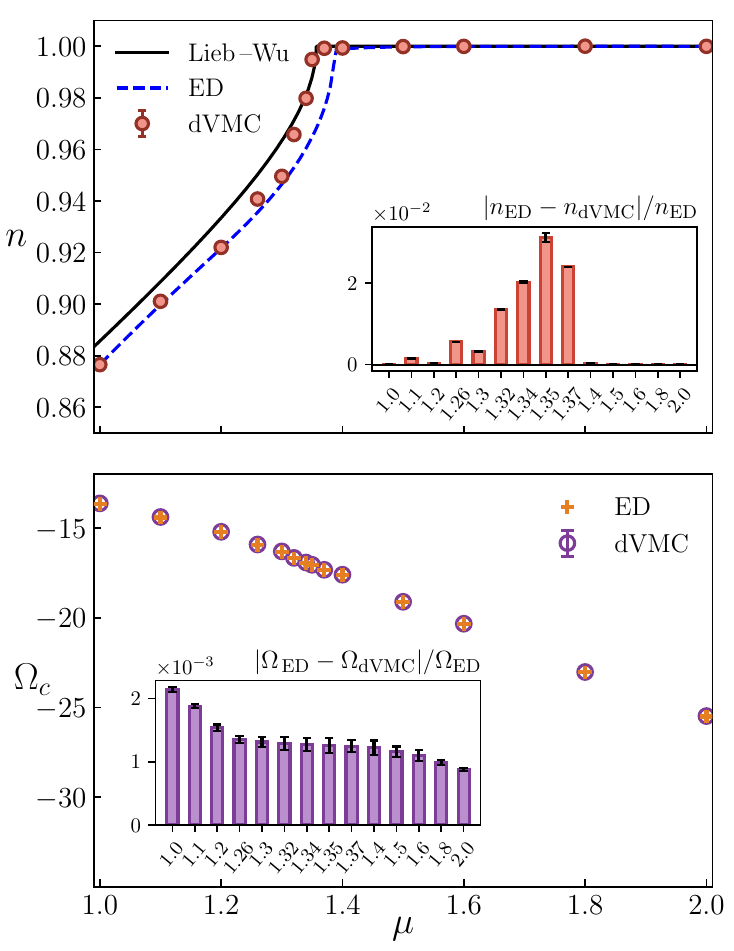}
    \end{center}
        \caption{(top) Average density versus chemical potential computed with dVMC and ED
    impurity solvers. The exact result from Lieb \& Wu \cite{Lieb-Wu} is given by the black curve.
    (bottom) Grand potential (units of $t$) of the cluster versus chemical potential. 
    The purple circles show the CDMFT-dVMC result and the orange crosses show the
    ED result computed with the bath parameters from the converged CDMFT-dVMC
    iterations.
    The error bars represent the error of the CDMFT-dVMC result (see
    Appendix~\ref{App:vs_Nb} for details). Note that the same error is given in the inset as a fraction
    of the ED result.
    The system has $N_s$ = 8 cluster sites, $N_b = 4$ bath sites and $N_e = 12$ electrons.
    \label{fig:n_E_vs_mu_dVMC_vs_ED}}
\end{figure}

Having calibrated the accuracy of the dVMC solver on small clusters, we proceed with a set of calculations on
intermediate sized clusters. In the top panel of Fig.~\ref{fig:n_E_vs_mu_dVMC_vs_ED} we again show the average lattice density versus 
chemical potential, computed with the dVMC and exact diagonalization impurity solvers. We find that the dVMC results are in 
good agreement with the exact diagonalization results in both the metallic and insulating limits, for values of $\mu < 1.3$ and $\mu > 1.4$.
In the region near the transition between the metallic and insulating states, $\mu \in [1.32,1.38]$, the dVMC results are in qualitative 
agreement with the exact diagonalization results, however, the dVMC impurity solver obtains a different value of the critical chemical 
potential $\mu_c$, which indicates the edge of the gap. The dVMC curve still shows a sharp increase in slope as the chemical
potential approaches $\mu_c$, but this increase occurs at a slightly lower value of chemical potential than the corresponding
exact diagonalization result. 

In order to better understand the behavior of the dVMC solver, we next consider the accuracy of the 
ground state of the cluster obtained by the solver. The bottom panel of Fig.~\ref{fig:n_E_vs_mu_dVMC_vs_ED} 
shows the grand potential of the cluster, $\Omega_c$, versus chemical potential. In the top panel, at 
each chemical potential, the dVMC value represents the average over the converged CDMFT iterations (see Appendix~\ref{App:vs_Nb}), 
whereas the ED result is the value at the final CDMFT iteration. In the bottom panel, the dVMC value is again obtained by
averaging over the converged iterations, whereas the ED value is computed using the bath parameters from the converged 
CDMFT-dVMC iterations. The grand potential of the cluster computed by dVMC is generally in excellent agreement with the exact 
diagonalization result, typically within $\sim 0.2\%$. The dVMC result tends to become more accurate as a function
of increasing chemical potential, with the most accurate results being obtained after the transition to the insulating state above
$\mu \sim 1.4$. We note however, that the error bars are largest in the region near $\mu_c$~\cite{Koch2008}, for $\mu \in [1.3,1.6]$.
This suggests that the ground state energy landscape in this region is complicated, which is reflected in the 
larger variance in the representation of the ground state achieved by the dVMC solver. The dVMC description of the ground state
is a subject we will touch upon in a later section. 

\begin{figure*}[!ht]
    \begin{center}
           \includegraphics[width=\textwidth]{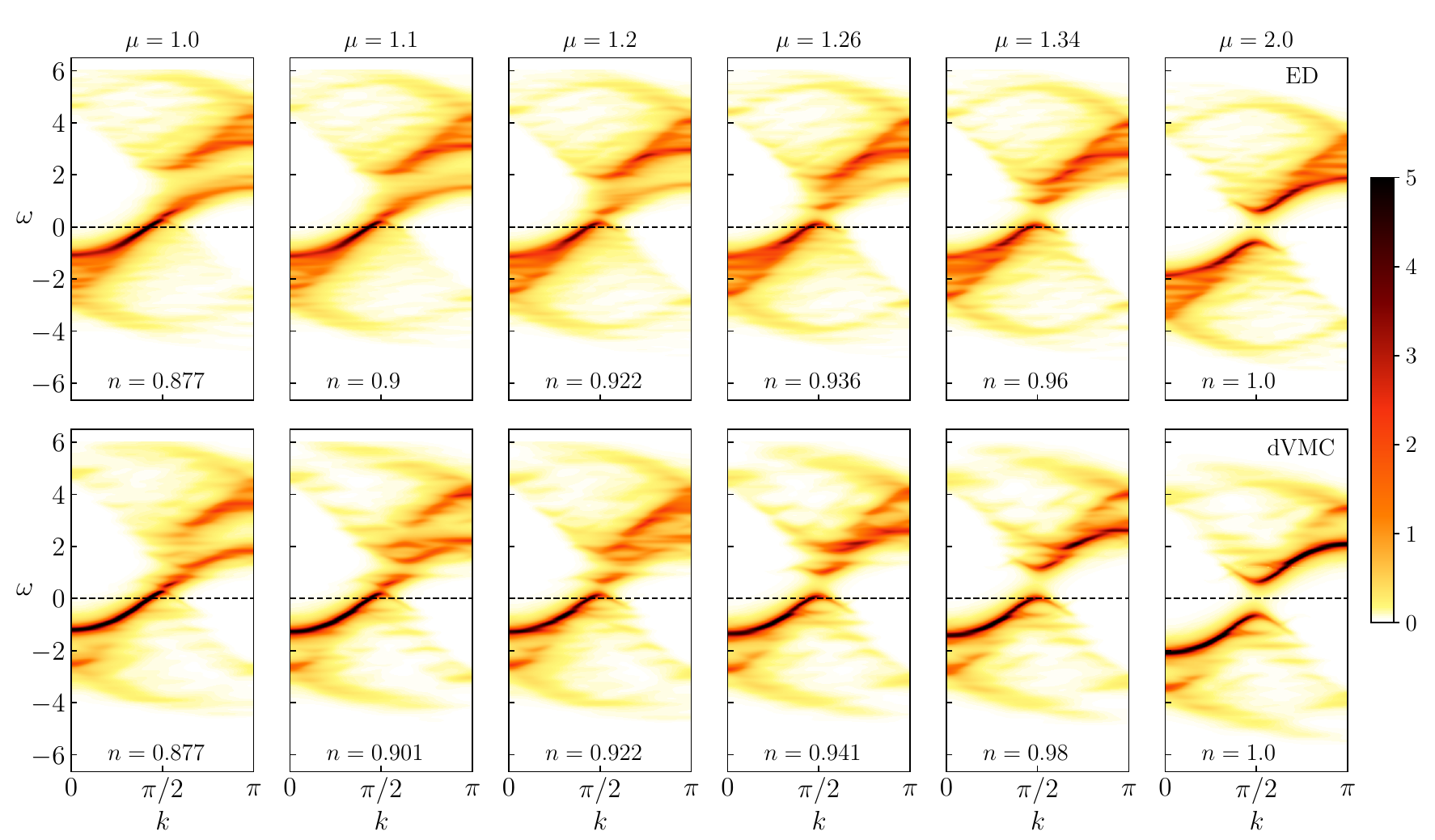}
    \end{center}
        \caption{$A(k,\omega)$ versus average density for $N_c = 8$ and $N_b = 4$ system from CDMFT-ED (top row) and CDMFT-dVMC (bottom row) systems. Each column corresponds to a specific $\mu$, with the resulting average lattice density given at the bottom of each plot.}
    \label{fig:spectra_vs_mu_L8_4b}
\end{figure*}

As highlighted above, the ability to compute dynamical properties of strongly-correlated systems is one of 
the central motivations behind the development of this technique. We therefore proceed by computing the
spectral function, $A(k,\omega)$, for several of the systems presented in Fig.~\ref{fig:n_E_vs_mu_dVMC_vs_ED}. 
To further gauge the accuracy of the dVMC impurity solver we compute the same quantity with both the dVMC and
exact diagonalization solvers. In the top row of Fig.~\ref{fig:spectra_vs_mu_L8_4b} we present the spectral function versus
average density obtained with the exact diagonalization impurity solver and in the bottom row we show
the result obtained with the dVMC solver.

We find that the CDMFT-dVMC results capture the same basic qualitative features as the CDMFT-ED results.
In both sets of calculations, a single band crosses the Fermi level at lower density, which gradually loses spectral weight 
until a gap opens that is symmetric about the Fermi level at half filling. 

Finally, to complement the results presented above, we compute the local density at
several values of chemical potential. As in earlier results, the CDMFT-dVMC values
are obtained by averaging over the set of converged iterations, whereas the CDMFT-ED values
are taken from the final CDMFT-ED iteration.

\begin{figure}[!ht]
    \begin{center}
           \includegraphics[width=\columnwidth]{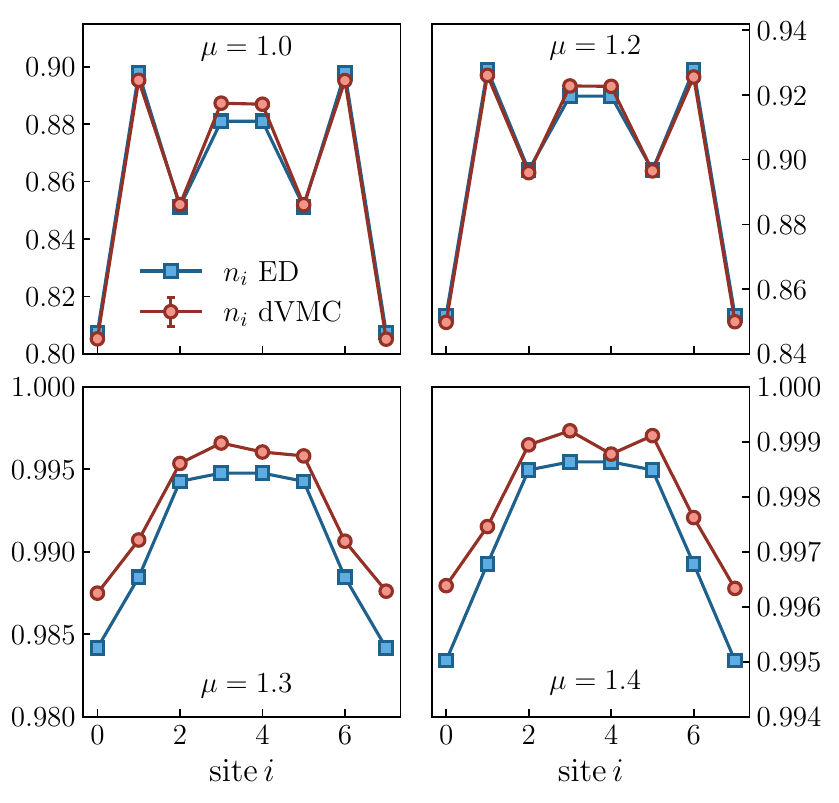}
    \end{center}
        \caption{Local density, $n_i$, versus cluster site, $i$, from CDMFT-dVMC and CDMFT-ED. 
        Each system has  $N_c$ = 8 cluster sites, $N_b = 4$ bath sites, and $N_e = 12$ electrons.}
    \label{fig:local_dens_L8_4b_dVMC_vs_ED}
\end{figure}

The CDMFT-dVMC results for the local density are generally in good quantitative agreement 
with the CDMFT-ED results (Fig.~\ref{fig:local_dens_L8_4b_dVMC_vs_ED}). At smaller values of chemical 
potential (top row) the CDMFT-ED result shows an oscillatory feature that the CDMFT-dVMC results match to
within $\sim 1\%$ error. At larger values of chemical potential (bottom row) the CDMFT-dVMC results are even 
more accurate, in this case within $\sim 0.25\%$ of the CDMFT-ED result.

\subsection{Results for large systems}

\begin{figure}[!ht]
    \begin{center}
           \includegraphics[width=\columnwidth]{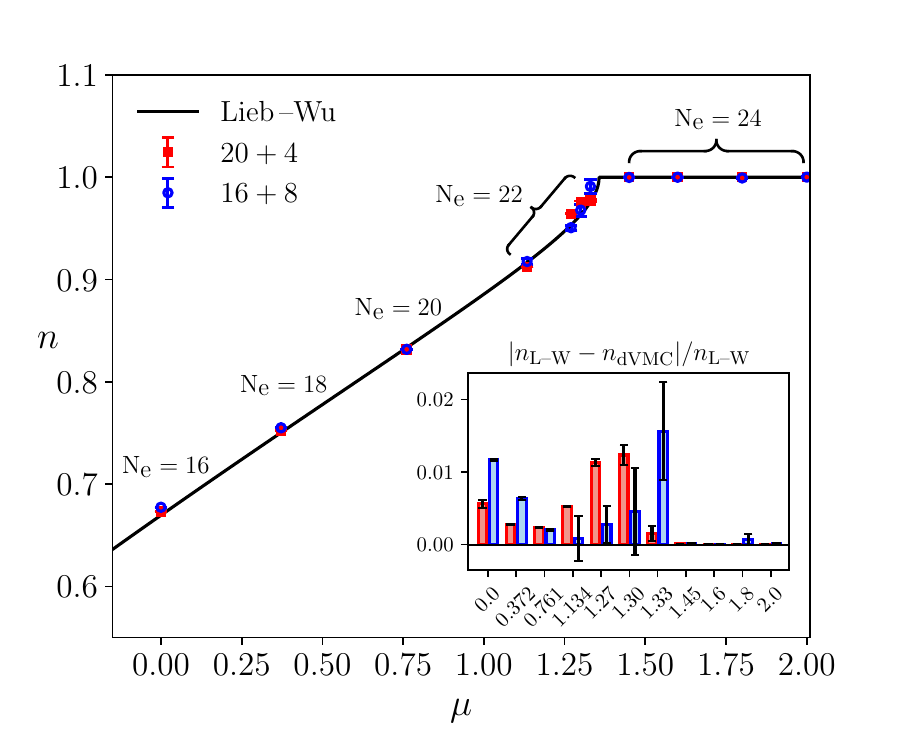}
    \end{center}
    \caption{Average density versus chemical potential for systems with 24 total sites. The exact Lieb\,--Wu result
    is given by the black curve. (inset) Relative error
    between Lieb\,--Wu result and CDMFT result with dVMC impurity solver at each value of $\mu$ (given by $x$-axis)
    studied. The error bars represent the error of the CDMFT-dVMC result (see
     Appendix~\ref{App:vs_Nb} for details). Note that the same error is given in the inset as a fraction
    of the ED result.
    The number of electrons, $N_e$,
     for each calculation is given above the corresponding data point or set of data points.  \label{n_vs_mu_large_systems}
}
\end{figure}

In the preceding sections we presented a thorough set of benchmarks on small and intermediate
sized clusters to establish the accuracy of the dVMC impurity solver within CDMFT. Here we perform
a set of calculations on large clusters, beyond the reach of exact diagonalization solvers. The ability
to treat larger clusters reduces finite size effects and provides a clearer understanding of the behavior of the 
method as a function of the number of cluster sites as well as bath sites. Our results for the average
lattice density versus chemical potential are summarized in Fig.~\ref{n_vs_mu_large_systems}.

We used two different clusters for this set of calculations, both with 24 total sites. We observe that
for small to intermediate chemical potential ($\mu \leq 1.2$) the 20 site cluster with 4 bath sites ($20+4$) and
the 16 site cluster with 8 bath sites ($16+8$) both obtain results for the average density in good agreement with
the Lieb\,--Wu result, generally within $\sim 1\%$. Similarly to the results for smaller clusters, the
agreement between the CDMFT-dVMC result and the Lieb\,--Wu result is somewhat worse towards the transition
to the insulating state. The Monte Carlo estimate of the error is also larger in this region, suggesting a potentially
complicated ground state landscape that poses a challenge for the dVMC approach, which is consistent with the 
previous benchmark against ED shown in Fig.~\ref{fig:n_E_vs_mu_dVMC_vs_ED}. For chemical potentials
beyond the transition, the CDMFT-dVMC results are exceptionally accurate.

\begin{figure*}[!ht]
    \begin{center}
           \includegraphics[width=\textwidth]{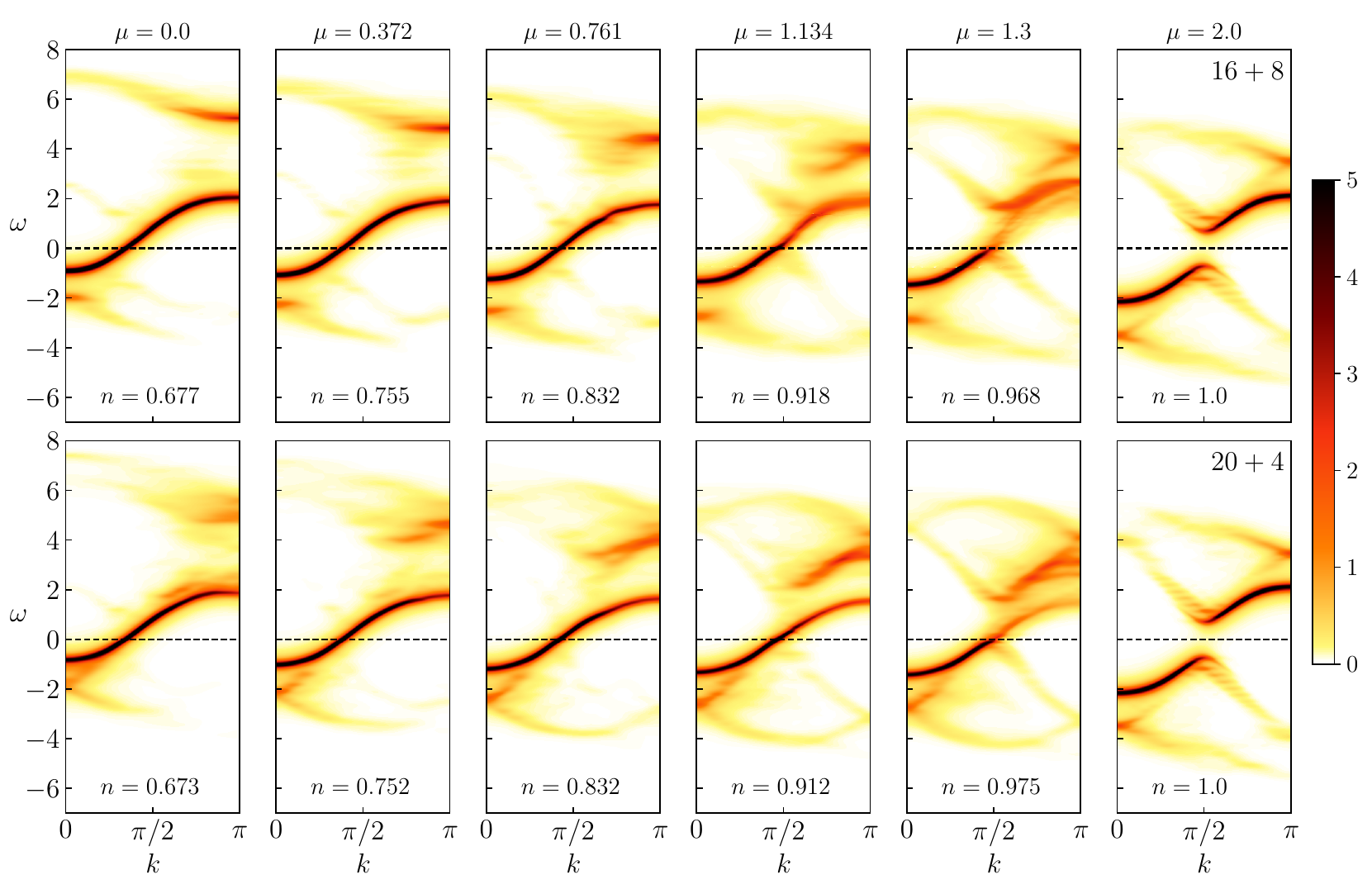}
    \end{center}
    \caption{$A(k,\omega)$ versus average density for $N_c = 16$ and $N_b = 8$ (top row) and $N_c = 20$ and $N_b = 4$ (bottom row) systems from CDMFT-dVMC. Each column corresponds to
    a specific $\mu$, with the resulting average lattice density given at the bottom of each plot. \label{spectra_vs_mu_large_systems}
}
\end{figure*}

In Fig.~\ref{spectra_vs_mu_large_systems} we show the spectral function versus average density for
the $16+8$ system (top row) and the $20+4$ system (bottom row). The results for both clusters reliably capture the major qualitative
features of the physics at increasing values of average density \cite{Kohno2010}. At low density there is a single band with
significant spectral weight crossing the Fermi level. As the density increases, this band loses spectral
weight above the Fermi level and eventually a gap opens, with the Fermi level lying in the middle of
the gap at the particle-hole symmetric point ($\mu=2.0$ in this case). The differences between the results presented in the top row of 
Fig.~\ref{spectra_vs_mu_large_systems} and the bottom row provide a good estimate of the error or imprecision of the 
method. While the average densities are not identical for the two systems, they capture the same physics and qualitative 
behavior of the spectral function. Importantly, despite the small discrepancies between these two sets of results, the 
quality of the spectra is considerably improved, and less discretized, than the result for smaller clusters, from ED or dVMC (Fig.~\ref{fig:spectra_vs_mu_L8_4b}), 
which shows visible finite size effects.

\begin{figure}[!ht]
    \begin{center}
           \includegraphics[width=0.9\columnwidth]{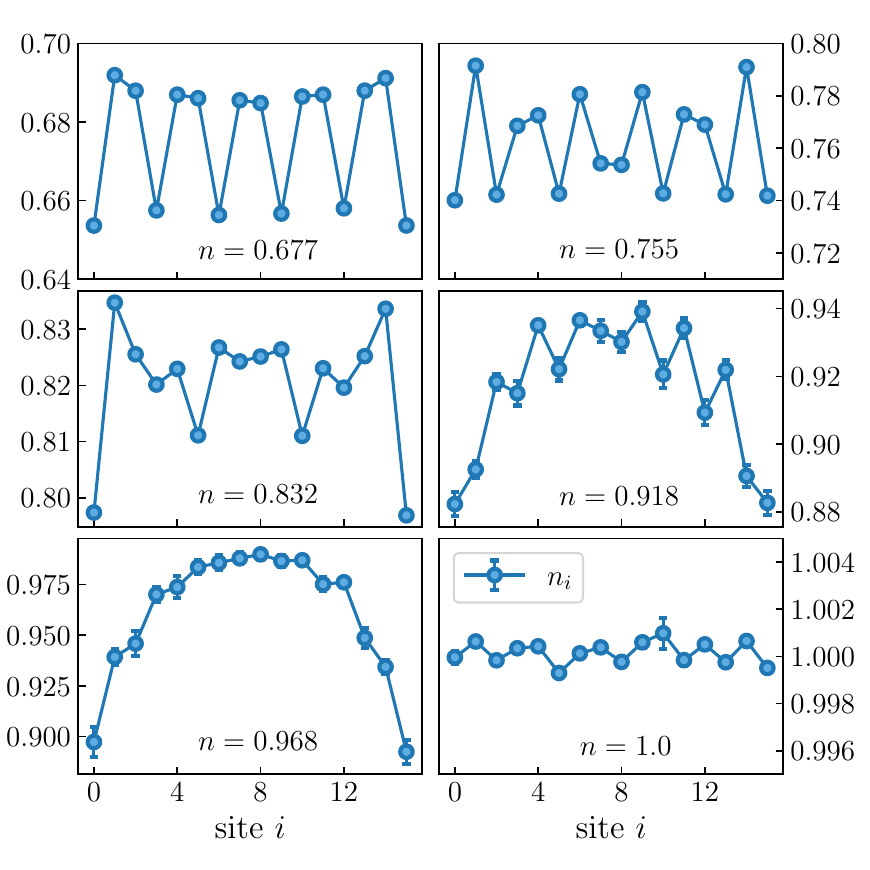}
    \end{center}
    \caption{Local density, $n_i$, versus cluster site, $i$, at various values of total average density for a $16+8$ system. 
    An oscillation in the local density, likely related to the finite cluster size and average density, is evident at $n=0.677$.
    This oscillation disappears at larger values of average density until the local density is essentially constant at
    $n=1.0$.\label{fig:local_dens_L16-8b}
}
\end{figure}

\begin{figure}[!ht]
    \begin{center}
           \includegraphics[width=0.9\columnwidth]{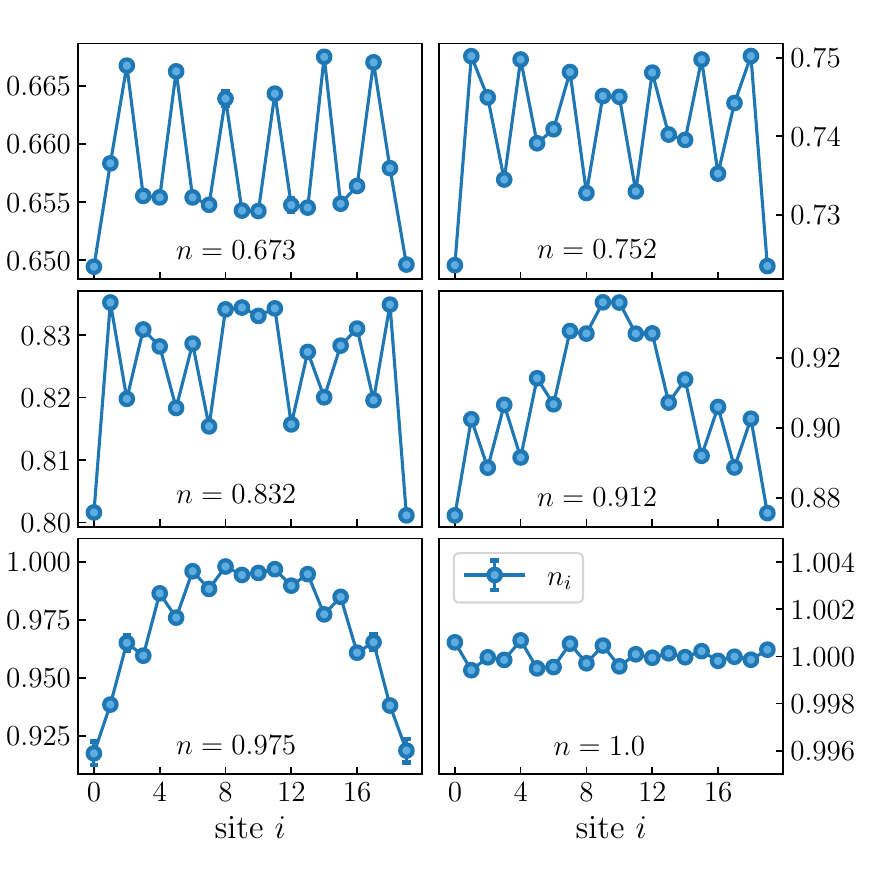}
    \end{center}
    \caption{Local density, $n_i$, versus cluster site, $i$, at various values of total average density for a $20+4$ system.
    A similar oscillation to the $16+8$ system is evident at $n=0.673$, though with a different period, suggesting
    that it is indeed a finite-size effect. The local density again becomes more uniform as the average density approaches 1.0.\label{fig:local_dens_L20-4b}
}
\end{figure}

Finally, we compute the local density at several values of average total lattice density for
both clusters. Far from half-filling (upper left panel of Figs.~\ref{fig:local_dens_L16-8b}, \ref{fig:local_dens_L20-4b}),
the local density shows a modulation that is likely induced by the finite size of the lattice in combination
with the particular value of total average lattice density. No charge order is observed in the Lieb\,--Wu solution. 
As the total density increases, this oscillation disappears and the local density becomes more uniform towards 
the center of the cluster. At half-filling the local density is essentially constant, as expected. 

\section{Discussion}
\label{se::discussion}

In this section we note some of the remaining technical challenges that currently limit the performance of
the technique, and suggest potential improvements reserved for future work. 

One of the limitations of the method is the description of the ground state. In the results presented in 
Fig.~\ref{fig:n_E_vs_mu_dVMC_vs_ED}, we observe that the
error in the grand potential of the cluster relative to the ED result is generally quite small, on the order of
$0.1$--\,$0.2\%$, however, the error bar grows in the region near the transition between the metallic
and insulating states. This behavior suggests that there may be closely spaced local minima in the
ground state energy landscape, which leads to a larger variance in the dVMC result. 
The error in the average lattice density is also largest near the transition, indicating that some
of the error in the lattice Green function may come from inaccuracies, or the higher variance, of
the ground state. 

One means of addressing this limitation is to consider different, more flexible, variational ground state ansatzes,
that might provide a more accurate description of the ground state~\cite{Wu_Rossi_Vicentini_Astrakhantsev_Becca_Cao_Carrasquilla_Ferrari_Georges_Hibat-Allah}. 
The past several years have seen
considerable progress in this direction, with the development of innovative variational Monte Carlo
approaches, including many inspired by ideas from machine learning \cite{Han2019,Choo2020,Pfau2020,Stokes2020} or tensor networks~\cite{Kloss_Thoenniss_Sonner_Lerose_Fishman_Stoudenmire_Parcollet_Georges_Abanin_2023}. The application of these
approaches within dVMC remains an interesting prospect, with the potential to produce higher
accuracy results for dynamical properties.

One other potential source of error in the technique is the choice of excitations, i.e. the choice of the non-orthogonal basis used to express the one particle excited sectors of the Hamiltonian. As illustrated in
Fig.~\ref{fig:n_E_vs_mu_dVMC_vs_ED}, the overall error in the average lattice density seems
to depend somewhat on the accuracy of the ground state, but the ground state generally agrees
quite well with the ED result. While it is difficult to disentangle the sources of error, given that the CDMFT-dVMC
technique involves a self-consistency procedure comprising multiple variational minimizations,
it may be possible to reduce the overall error by further optimizing the choice of excitations, as discussed in Appendix~\ref{App:vs_Nexc}.

\section{Conclusion}

The dVMC technique has proven to be an accurate method to compute the Green function
for models of strongly-correlated electrons, with and without periodic boundary conditions and
translational invariance \cite{Charlebois2020,Rosenberg2022}. These previous developments
laid the foundation for the approach to be implemented as an impurity solver in various quantum
cluster techniques. Here we have focused on CDMFT, which includes the additional components
of coupling to bath sites and the self-consistent optimization of the bath parameters. 

We have introduced an impurity solver for CDMFT based on the dVMC technique and performed
a systematic set of benchmarks on the 1D Hubbard model. We compare against CDMFT-ED results on smaller clusters 
and against the Lieb\,--Wu solution for larger clusters. As we have shown, the approach is capable of 
achieving impressive accuracy (generally within $1.5\%$ error), and importantly, scales reasonably with system size, which 
makes it possible to treat large systems.

Though we have focused here on the 1D Hubbard model, the approach can be applied to a wide range of strongly correlated 
Hamiltonians, including two- and three-dimensional systems. Another goal of future work will be to extend the technique to treat superconducting systems, which requires 
measurement of the Nambu Green function. The ability to treat larger clusters is an important means of improving the approximation 
underlying quantum cluster methods. The approach we have introduced here extends the range of these already powerful methods, 
and holds the promise of new insights on the physics of strongly-correlated electrons.

\begin{acknowledgments}

P.R. was supported by a postdoctoral fellowship from the Canada First Research Excellence Fund through Institut quantique.
Computing resources were provided by the Digital Research Alliance of Canada (formerly Compute Canada) and Calcul Qu\'ebec. 

\end{acknowledgments}

\appendix

\section{Convergence versus number of bath sites}
\label{App:vs_Nb}

\begin{figure}[!ht]
    \begin{center}
           \includegraphics[width=\columnwidth]{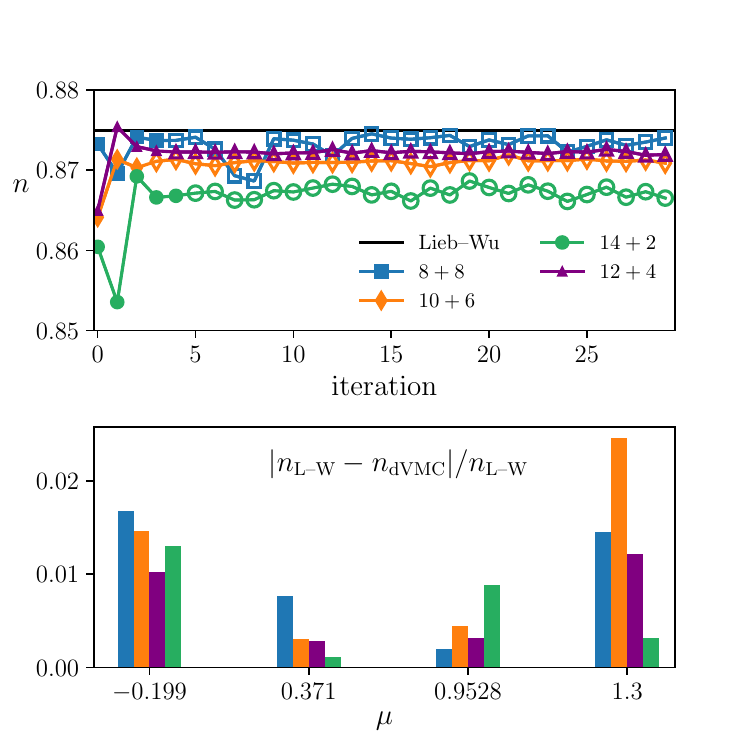}
    \end{center}
    \caption{(top) CDMFT-dVMC measurements of the average density versus iteration at chemical potential $\mu=0.9528$. The open symbols
    indicate iterations that are included in the computation of the converged value of the average density. (bottom) Relative error with respect to the 
    Lieb\,--Wu result at several values of chemical potential. \label{fig:conv_vs_Nbath}}
 \end{figure}

In Fig.~\ref{fig:conv_vs_Nbath} we show several calculations of
the average density for a system with a total of 16 sites, but different numbers of bath sites (See also Ref.\onlinecite{Koch2008}).
In the upper panel we show the average lattice density computed at each iteration of
the CDMFT self-consistency loop for each system. As alluded to in the main text, in 
the CDMFT-dVMC approach we determine convergence by inspection of the bath parameters.
Once the calculation has converged we perform an additional set of iterations. The average
value of an observable and the associated error bar are obtained by computing the average and
standard deviation of the measurements of that observable over this set of converged iterations.

In the upper panel of Fig.~\ref{fig:conv_vs_Nbath} the converged iterations are indicated by
the open symbols. For the chemical potential shown in this panel, we observe that the result 
for the system with the largest number of bath sites is closest to the exact result. However,
this behavior is not consistent across all the values of chemical potential we have studied,
as illustrated in the lower panel. For instance, while the $8+8$ system is the most accurate at $\mu=0.9528$,
it is the least accurate at $\mu = -0.199$.   

\section{Convergence versus number of excitations}
 \label{App:vs_Nexc}
 
\begin{figure*}[!ht]
    \begin{center}
           \includegraphics[width=0.9\textwidth]{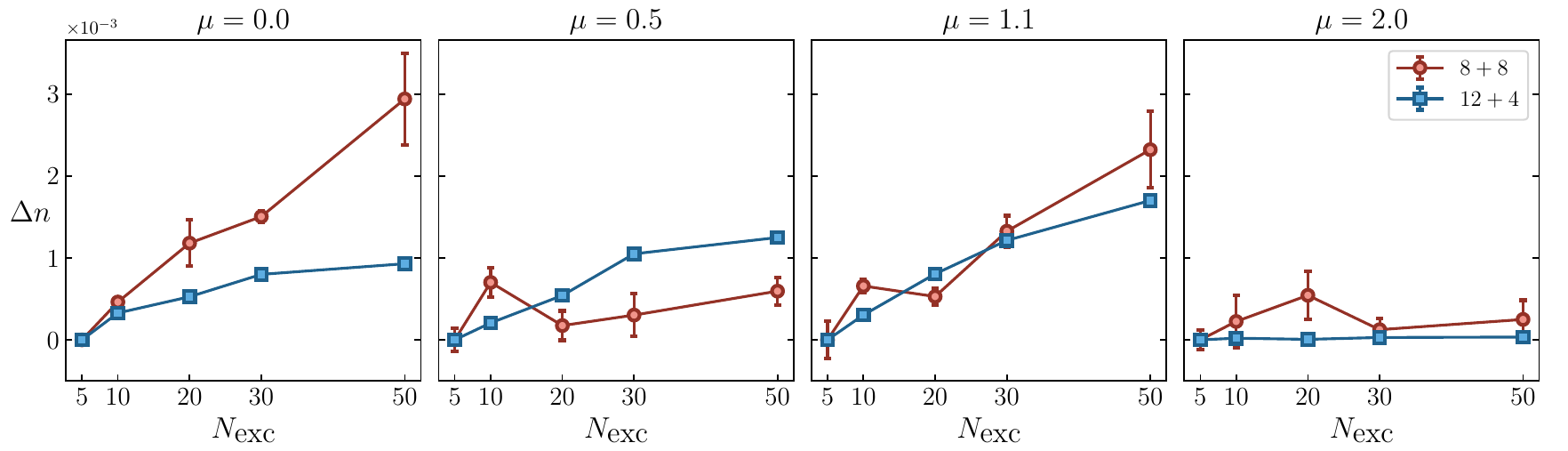}
    \end{center}
    \caption{Convergence versus $N_\textrm{exc}$. Each panel plots $\Delta n \equiv \vert n(N_\textrm{exc})-n(5) \vert$, which gives the change in the average lattice density as a function the number of excitations, relative to the value at $N_\textrm{exc}=5$.
    Note, we refer to the number of excitations included before filtering. \label{fig:conv_vs_Nexc}}
 \end{figure*}

In this Appendix we study the behavior of the technique as a function of the number
of excitations. In Fig.~\ref{fig:conv_vs_Nexc} we show several calculations of 
the average lattice density for a $12+4$ and an $8+8$ system across a range of $N_\textrm{exc}$,
for several values of chemical potential.

We observe that the result for the average lattice density shows relatively little dependence on
the number of excitations included before filtering, as the difference between the result with 
$N_\textrm{exc}=5$ and $N_\textrm{exc}=50$ is below $\sim5\times10^{-3}$, or $~0.5\%$ of the
final result for the average lattice density, at all values of chemical potential studied. We observe
that the dependence is particularly small for systems within the insulating state (rightmost panel). 
This is true for both the $12+4$ and the $8+8$ system, however, the $8+8$ system has larger error bars, likely
due in part to the larger number of variational parameters.

%

\end{document}